\documentclass[aps,prl,reprint,superscriptaddress,showpacs,showkeys]{revtex4-1}

\usepackage{graphicx}
\usepackage{natbib}

\begin{document}


\title{First results on low-mass WIMP from the CDEX-1 experiment \\at the China Jinping underground Laboratory}


\affiliation{Key Laboratory of Particle and Radiation Imaging (Ministry of Education) and Department of Engineering Physics, Tsinghua University, Beijing 100084}
\affiliation{Department of Nuclear Physics, China Institute of Atomic Energy, Beijing 102413}
\affiliation{School of Physics, Nankai University, Tianjin 300071}
\affiliation{NUCTECH Company, Beijing 10084}
\affiliation{School of Physical Science and Technology, Sichuan University, Chengdu 610065}
\affiliation{YaLong River Hydropower Development Company, Chengdu 610051}
\affiliation{Institute of Physics, Academia Sinica, Taipei 11529}
\affiliation{Department of Physics, Banaras Hindu University, Varanasi 221005}

\author{W. Zhao}
\affiliation{Key Laboratory of Particle and Radiation Imaging (Ministry of Education) and Department of Engineering Physics, Tsinghua University, Beijing 100084}
\author{Q. Yue}
\email{Corresponding author: yueq@mail.tsinghua.edu.cn}
\affiliation{Key Laboratory of Particle and Radiation Imaging (Ministry of Education) and Department of Engineering Physics, Tsinghua University, Beijing 100084}
\author{K.J. Kang}
\affiliation{Key Laboratory of Particle and Radiation Imaging (Ministry of Education) and Department of Engineering Physics, Tsinghua University, Beijing 100084}
\author{J.P. Cheng}
\affiliation{Key Laboratory of Particle and Radiation Imaging (Ministry of Education) and Department of Engineering Physics, Tsinghua University, Beijing 100084}
\author{Y.J. Li}
\affiliation{Key Laboratory of Particle and Radiation Imaging (Ministry of Education) and Department of Engineering Physics, Tsinghua University, Beijing 100084}
\author{S.T. Lin}
\altaffiliation{Participating as a member of TEXONO Collaboration}
\affiliation{Institute of Physics, Academia Sinica, Taipei 11529}
\author{Y. Bai}
\affiliation{School of Physics, Nankai University, Tianjin 300071}
\author{Y. Bi}
\affiliation{School of Physical Science and Technology, Sichuan University, Chengdu 610065}
\author{J.P. Chang}
\affiliation{NUCTECH Company, Beijing 10084}
\author{N. Chen}
\affiliation{Key Laboratory of Particle and Radiation Imaging (Ministry of Education) and Department of Engineering Physics, Tsinghua University, Beijing 100084}
\author{N. Chen}
\affiliation{Key Laboratory of Particle and Radiation Imaging (Ministry of Education) and Department of Engineering Physics, Tsinghua University, Beijing 100084}
\author{Q.H. Chen}
\affiliation{Key Laboratory of Particle and Radiation Imaging (Ministry of Education) and Department of Engineering Physics, Tsinghua University, Beijing 100084}
\author{Y.H. Chen}
\affiliation{YaLong River Hydropower Development Company, Chengdu 610051}
\author{Y.C. Chuang}
\altaffiliation{Participating as a member of TEXONO Collaboration}
\affiliation{Institute of Physics, Academia Sinica, Taipei 11529}
\author{Z. Deng}
\affiliation{Key Laboratory of Particle and Radiation Imaging (Ministry of Education) and Department of Engineering Physics, Tsinghua University, Beijing 100084}
\author{C. Du}
\affiliation{Key Laboratory of Particle and Radiation Imaging (Ministry of Education) and Department of Engineering Physics, Tsinghua University, Beijing 100084}
\author{Q. Du}
\affiliation{Key Laboratory of Particle and Radiation Imaging (Ministry of Education) and Department of Engineering Physics, Tsinghua University, Beijing 100084}
\author{H. Gong}
\affiliation{Key Laboratory of Particle and Radiation Imaging (Ministry of Education) and Department of Engineering Physics, Tsinghua University, Beijing 100084}
\author{X.Q. Hao}
\affiliation{Key Laboratory of Particle and Radiation Imaging (Ministry of Education) and Department of Engineering Physics, Tsinghua University, Beijing 100084}
\author{H.J. He}
\affiliation{Key Laboratory of Particle and Radiation Imaging (Ministry of Education) and Department of Engineering Physics, Tsinghua University, Beijing 100084}
\author{Q.J. He}
\affiliation{Key Laboratory of Particle and Radiation Imaging (Ministry of Education) and Department of Engineering Physics, Tsinghua University, Beijing 100084}
\author{X.H. Hu}
\affiliation{School of Physics, Nankai University, Tianjin 300071}
\author{H.X. Huang}
\affiliation{Department of Nuclear Physics, China Institute of Atomic Energy, Beijing 102413}
\author{T.R. Huang}
\altaffiliation{Participating as a member of TEXONO Collaboration}
\affiliation{Institute of Physics, Academia Sinica, Taipei 11529}
\author{H. Jiang}
\affiliation{Key Laboratory of Particle and Radiation Imaging (Ministry of Education) and Department of Engineering Physics, Tsinghua University, Beijing 100084}
\author{H.B. Li}
\altaffiliation{Participating as a member of TEXONO Collaboration}
\affiliation{Institute of Physics, Academia Sinica, Taipei 11529}
\author{J.M. Li}
\affiliation{Key Laboratory of Particle and Radiation Imaging (Ministry of Education) and Department of Engineering Physics, Tsinghua University, Beijing 100084}
\author{J. Li}
\affiliation{Key Laboratory of Particle and Radiation Imaging (Ministry of Education) and Department of Engineering Physics, Tsinghua University, Beijing 100084}
\author{J. Li}
\affiliation{NUCTECH Company, Beijing 10084}
\author{X. Li}
\affiliation{Department of Nuclear Physics, China Institute of Atomic Energy, Beijing 102413}
\author{X.Y. Li}
\affiliation{School of Physics, Nankai University, Tianjin 300071}
\author{Y.L. Li}
\affiliation{Key Laboratory of Particle and Radiation Imaging (Ministry of Education) and Department of Engineering Physics, Tsinghua University, Beijing 100084}
\author{H.Y. Liao}
\altaffiliation{Participating as a member of TEXONO Collaboration}
\affiliation{Institute of Physics, Academia Sinica, Taipei 11529}
\author{F.K. Lin}
\altaffiliation{Participating as a member of TEXONO Collaboration}
\affiliation{Institute of Physics, Academia Sinica, Taipei 11529}
\author{S.K. Liu}
\affiliation{School of Physical Science and Technology, Sichuan University, Chengdu 610065}
\author{L.C. L\"{u}}
\affiliation{Key Laboratory of Particle and Radiation Imaging (Ministry of Education) and Department of Engineering Physics, Tsinghua University, Beijing 100084}
\author{H. Ma}
\affiliation{Key Laboratory of Particle and Radiation Imaging (Ministry of Education) and Department of Engineering Physics, Tsinghua University, Beijing 100084}
\author{S.J. Mao}
\affiliation{NUCTECH Company, Beijing 10084}
\author{J.Q. Qin}
\affiliation{Key Laboratory of Particle and Radiation Imaging (Ministry of Education) and Department of Engineering Physics, Tsinghua University, Beijing 100084}
\author{J. Ren}
\affiliation{Department of Nuclear Physics, China Institute of Atomic Energy, Beijing 102413}
\author{J. Ren}
\affiliation{Key Laboratory of Particle and Radiation Imaging (Ministry of Education) and Department of Engineering Physics, Tsinghua University, Beijing 100084}
\author{X.C. Ruan}
\affiliation{Department of Nuclear Physics, China Institute of Atomic Energy, Beijing 102413}
\author{M.B. Shen}
\affiliation{YaLong River Hydropower Development Company, Chengdu 610051}
\author{L. Singh}
\altaffiliation{Participating as a member of TEXONO Collaboration}
\affiliation{Institute of Physics, Academia Sinica, Taipei 11529}
\affiliation{Department of Physics, Banaras Hindu University, Varanasi 221005}
\author{M.K. Singh}
\altaffiliation{Participating as a member of TEXONO Collaboration}
\affiliation{Institute of Physics, Academia Sinica, Taipei 11529}
\affiliation{Department of Physics, Banaras Hindu University, Varanasi 221005}
\author{A.K. Soma}
\altaffiliation{Participating as a member of TEXONO Collaboration}
\affiliation{Institute of Physics, Academia Sinica, Taipei 11529}
\affiliation{Department of Physics, Banaras Hindu University, Varanasi 221005}
\author{J. Su}
\affiliation{Key Laboratory of Particle and Radiation Imaging (Ministry of Education) and Department of Engineering Physics, Tsinghua University, Beijing 100084}
\author{C.J. Tang}
\affiliation{School of Physical Science and Technology, Sichuan University, Chengdu 610065}
\author{C.H. Tseng}
\altaffiliation{Participating as a member of TEXONO Collaboration}
\affiliation{Institute of Physics, Academia Sinica, Taipei 11529}
\author{J.M. Wang}
\affiliation{YaLong River Hydropower Development Company, Chengdu 610051}
\author{L. Wang}
\affiliation{School of Physical Science and Technology, Sichuan University, Chengdu 610065}
\author{Q. Wang}
\affiliation{Key Laboratory of Particle and Radiation Imaging (Ministry of Education) and Department of Engineering Physics, Tsinghua University, Beijing 100084}
\author{H.T. Wong}
\altaffiliation{Participating as a member of TEXONO Collaboration}
\affiliation{Institute of Physics, Academia Sinica, Taipei 11529}
\author{S.Y. Wu}
\affiliation{YaLong River Hydropower Development Company, Chengdu 610051}
\author{W. Wu}
\affiliation{School of Physics, Nankai University, Tianjin 300071}
\author{Y.C. Wu}
\affiliation{Key Laboratory of Particle and Radiation Imaging (Ministry of Education) and Department of Engineering Physics, Tsinghua University, Beijing 100084}
\author{Y.C. Wu}
\affiliation{NUCTECH Company, Beijing 10084}
\author{Z.Z. Xianyu}
\affiliation{Key Laboratory of Particle and Radiation Imaging (Ministry of Education) and Department of Engineering Physics, Tsinghua University, Beijing 100084}
\author{H.Y. Xing}
\affiliation{School of Physical Science and Technology, Sichuan University, Chengdu 610065}
\author{Y. Xu}
\affiliation{School of Physics, Nankai University, Tianjin 300071}
\author{X.J. Xu}
\affiliation{Key Laboratory of Particle and Radiation Imaging (Ministry of Education) and Department of Engineering Physics, Tsinghua University, Beijing 100084}
\author{T. Xue}
\affiliation{Key Laboratory of Particle and Radiation Imaging (Ministry of Education) and Department of Engineering Physics, Tsinghua University, Beijing 100084}
\author{L.T. Yang}
\affiliation{Key Laboratory of Particle and Radiation Imaging (Ministry of Education) and Department of Engineering Physics, Tsinghua University, Beijing 100084}
\author{S.W. Yang}
\altaffiliation{Participating as a member of TEXONO Collaboration}
\affiliation{Institute of Physics, Academia Sinica, Taipei 11529}
\author{N. Yi}
\affiliation{Key Laboratory of Particle and Radiation Imaging (Ministry of Education) and Department of Engineering Physics, Tsinghua University, Beijing 100084}
\author{C.X. Yu}
\affiliation{School of Physics, Nankai University, Tianjin 300071}
\author{H. Yu}
\affiliation{Key Laboratory of Particle and Radiation Imaging (Ministry of Education) and Department of Engineering Physics, Tsinghua University, Beijing 100084}
\author{X.Z. Yu}
\affiliation{School of Physical Science and Technology, Sichuan University, Chengdu 610065}
\author{X.H. Zeng}
\affiliation{YaLong River Hydropower Development Company, Chengdu 610051}
\author{Z. Zeng}
\affiliation{Key Laboratory of Particle and Radiation Imaging (Ministry of Education) and Department of Engineering Physics, Tsinghua University, Beijing 100084}
\author{L. Zhang}
\affiliation{NUCTECH Company, Beijing 10084}
\author{Y.H. Zhang}
\affiliation{YaLong River Hydropower Development Company, Chengdu 610051}
\author{M.G. Zhao}
\affiliation{School of Physics, Nankai University, Tianjin 300071}
\author{S.N. Zhong}
\affiliation{School of Physics, Nankai University, Tianjin 300071}
\author{Z.Y. Zhou}
\affiliation{Department of Nuclear Physics, China Institute of Atomic Energy, Beijing 102413}
\author{J.J. Zhu}
\affiliation{School of Physical Science and Technology, Sichuan University, Chengdu 610065}
\author{W.B. Zhu}
\affiliation{NUCTECH Company, Beijing 10084}
\author{X.Z. Zhu}
\affiliation{Key Laboratory of Particle and Radiation Imaging (Ministry of Education) and Department of Engineering Physics, Tsinghua University, Beijing 100084}
\author{Z.H. Zhu}
\affiliation{YaLong River Hydropower Development Company, Chengdu 610051}

\collaboration{CDEX Collaboration}
\noaffiliation



\date{\today}

\begin{abstract}
  The China Dark matter Experiment collaboration reports the first experimental limit on
  WIMP dark matter from 14.6 $\text{kg}$-$\text{day}$ of data taken with a 994 g p-type point-contact
  germanium detector at the China Jinping underground Laboratory where the rock
  overburden is more than 2400 m. The energy threshold achieved was 400 eVee. According
  to the 14.6 $\text{kg}$-$\text{day}$ live data, we placed  the limit of
  $\sigma_{\chi N} = 1.75 \times 10 ^{-40}$ cm$^2$ at 90\% confidence level on the
  spin-independent cross-section at WIMP mass of 7 GeV before differentiating
  bulk signals from the surface backgrounds.
\end{abstract}


\maketitle


There are many pieces of evidence from astroparticle physics and cosmology which
indicate that about one quarter of the mass of our universe is composed of dark matter
\cite{Beringer2012,*PlanckCollaboration2013a}. The nature of dark matter is unknown
except that it is coupled with matter via gravity. One of the possible candidates for
dark matter is the Weekly Interacting Massive Particles (WIMP, denoted by $\chi$ ) as motivated by many
new theories beyond the standard model \cite{Kelso2012}. Direct detection of WIMP
dark matter has been attempted with different detector technologies in the particle
physics domain \cite{Lewin1996}.

In recent years several experiments have expanded their coverage down to low
mass WIMP with $m_{\chi} < 10$ GeV \cite{Aalseth2008,*Aalseth2011,*Aalseth2011a,*Aalseth2012,
DAMACollaboration2011,*Bernabei2010,
Ahmed2010,*Ahmed2011,*CDMSCollaboration2012,*Agnese2013a,*Agnese2013b,
Angloher2012,
Aprile2011,*Angle2011,*Aprile2012,*Aprile2012a,
Lin2009,*TEXONOCollaboration2013}. A point-contact germanium detector can reach an energy threshold of
hundreds of eV while keeping almost the same energy resolution as the
traditional co-axial germanium detector \cite{Luke1989}. Thus it can be a
good choice for a low mass dark matter search. Based on our previous work \cite{Yue2004,*He2006,*LiXin2007},
the China Dark matter Experiment (CDEX) collaboration has formally started
a program aimed at the direct detection of low mass WIMP using a tonne-scale
germanium array detector system \cite{Kang2010b,*Yue2012}. As the first step, in
2011 the CDEX phase-I experiment (CDEX-1) started to test and run its first
prototype p-type point-contact germanium (PPCGe) detector with a crystal mass of
994 g. The experiment took place in China Jinping underground Laboratory (CJPL)
which was established at the end of 2010. With 2400 m rock overburden, CJPL is
the deepest operational underground laboratories for particle
physics in the world. The cosmic-ray flux in CJPL is down to 61.7
y$^{-1}$m$^{-2}$ \cite{WuYC2013}, and this makes it a very good site for
ultra-low background experiments such as dark matter search, double beta decay,
and so on.

The point-contact germanium detectors have also been used by several experiments
\cite{Aalseth2008,*Aalseth2011,*Aalseth2011a,*Aalseth2012, Lin2009,*TEXONOCollaboration2013}
to directly search for low mass WIMP. Due to the relative shallow cosmic-ray shielding, CoGeNT
and TEXONO used muon-veto detectors to decrease the direct and indirect background contribution from
cosmic-ray muons. The muon veto method can decrease the background contribution of the
cosmic-rays, but is less efficient to pick out the background from secondary prompt
and delayed cosmogenic radiation. The superior rock shielding of CJPL can minimize
the cosmogenic influence. However it is still important to study the background of
a prototype germanium detector to get more knowledge and experience
for evaluating and developing a tonne-scale experiment with germanium detector.

In this article, we report the first results from the CDEX-1 experiment. The
detailed information about the shielding system was described in reference
\cite{Kang2013}. This phase-I experiment did not install any active shielding
system such as an anti-Compton detector in order to understand and estimate the
background level of the PPCGe detector itself in CJPL, and it will helpful to compare
and learn the background contribution of the anti-Compton detector and its veto efficiency
installed in the next experiment phase. The passive shielding system was installed as,
from outside to inside, 1 m of polyethylene, 20 cm of lead, 20 cm of borated polyethylene
and 20 cm of oxygen-free high-conductivity copper.

\begin{figure}[h]
  \includegraphics[width=1.0\linewidth]{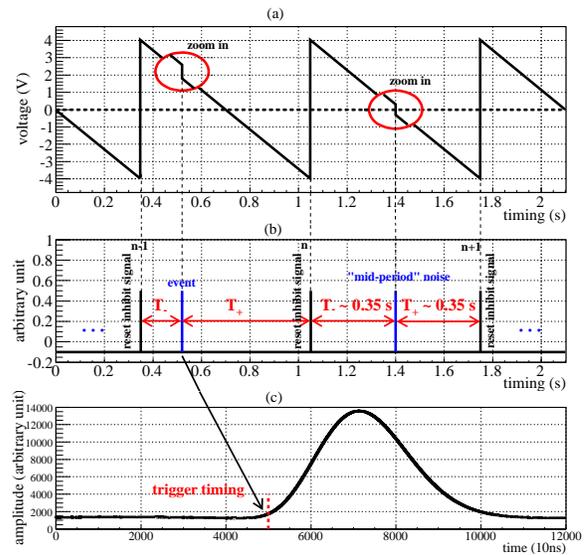}
  \caption{\label{fig:waveform} Output of the PPCGe detector showing the relative timing:
  (a) raw signals from reset preamplifier; (b) timing of reset-inhibit and typical physics
  and ``mid-period" noise events; (c) amplifier output with 12 $\mu$s shaping time.}
\end{figure}

Inside the cryostat of the PPCGe detector, the cylindrical germanium crystal has
an n$^+$ type contact on the outer surface and a tiny p$^+$ type contact as the
central electrode. A detailed illustration of the detector, electronics, and
data acquisition (DAQ) system as well as its basic performances was also described
in \cite{Kang2013}. Signals from the p$^+$ electrode of the PPCGe were
imported into a pulsed reset preamplifier which has 3
identical energy-related outputs. Two of them were loaded to shaping amplifiers at
6 $\mu$s (S$_{\text{p6}}$) and 12 $\mu$s (S$_{\text{p12}}$) shaping time, respectively.
The discriminator output of S$_{\text{p12}}$ supplied the trigger for the DAQ.
Another output was distributed to a timing amplifier (T$_{\text{p}}$) to
just amplify the pulse height of the raw-traces with few nano seconds shaping time.
The pulsed reset preamplifier worked in pulsed mode, so it was reset
to its initial condition by discharge the FET quickly marked by one reset inhibit signal.
The charge and discharge procedure of the preamplifier was shown in Fig.~\ref{fig:waveform}(a)
and the reset inhibit signals in Fig.~\ref{fig:waveform}(b). The reset inhibit signal was
rectangular and quasi-periodic, and the period was typically around 0.7 second which was related
to the leakage current of the PPCGe detector. Fig.~\ref{fig:waveform}(c) shows the output waveforms from
the main amplifier with 12 $\mu$s shaping time. The discrimination output of the reset inhibit signal
was also served as trigger to record the timing information of the discharging. Signals from
the n$^+$ electrode of the PPCGe were read out by a resistive feedback preamplifier, and then
were divided into 3 identical outputs followed by shaping amplifiers at 2 $\mu$s shaping time
but different gains. The S$_{\text{p6,12}}$, T$_{\text{p}}$ and n$^+$ electrode outputs
were sampled and recorded by a 100 MHz flash analog-to-digital convertor (FADC).
The discriminator outputs of the random trigger events (0.05Hz) from a pulse
generator were served as trigger and digitized as well.

A first-run 18.1 day was taken, and the DAQ dead time was 0.5\% as measured by
the random trigger events. We focus in this paper on the analysis of this data set.
Signals coming from S$_{\text{p12}}$ were chosen as the energy measurement. We defined
the pulse area of one event as its energy. The random trigger events were used for
zeroing calibration. Rectangular pulser signals with various amplitudes were injected
into the test input of the preamplifier of the PPCGe detector for trigger efficiency measurement.

We used two kinds of methods to do the data selection
and analysis. The first method is related to the timing information of one event.
The second one is related to the characteristics of the pulse shape of one event.

\begin{figure}[h]
  \includegraphics[width=1.0\linewidth]{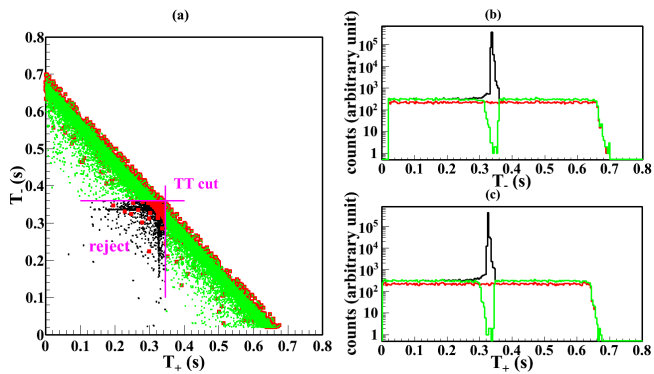}
  \caption{\label{fig:TT-cut} The scatter plots of T$_{\text{+}}$ and T$_{\text{-}}$ for
  random trigger events (red color) and background events before and after the TT cut (black and green color)
  have been shown in Fig.~\ref{fig:TT-cut}(a). The TT cut has also been overlaid on the scatter plot.
  The spectra before and after the TT cut have been given in the Fig.~\ref{fig:TT-cut}(b, c), at the
  same time, the T$_{\text{+}}$ and T$_{\text{-}}$ spectra for random trigger have also been shown in these figures. }
\end{figure}

In our experiment, noise events mostly occur in the middle range of the reset period while
the physics events and the random trigger events are uniformly distributed. This kind of ``mid-period"
noise was due to the interference of the output of the preamplifier and the zero level when they
are close to each other. As illustrated in Fig.~\ref{fig:TT-cut}(a) and (b), we denote T$_{\text{-}}$ as
the time interval between the event and its nearest prior-inhibit signal, and T$_{\text{+}}$ the time interval
between this event and its nearest post-inhibit signal. Fig.~\ref{fig:TT-cut}(a) shown the
relationship of T$_{\text{-}}$ and T$_{\text{+}}$ for both random trigger events and background events.
These can therefore be rejected by the ``TT Cut" which leads to a live time penalty.
The background spectra before and after the TT cut, as well the random trigger spectra,
have been shown in the Fig.~\ref{fig:TT-cut}(b, c).

The pulse shape discrimination is based on two methods.
The first method is the ``pedestals of S$_{\text{p6,12}}$ and T$_{\text{p}}$'' (Ped)
cuts which are applied to discriminate those events whose pedestals exhibit anomalous behaviour.
The criteria are derived from the random trigger events whose pedestal distributions
are considered consistent with the physical events. The signal efficiencies for the TT and Ped cuts,
which are independent on the pulse amplitude, were accurately measured with random trigger events,
and are 92.5 \% and 96.2 \%, respectively. A data set of 16.0 $\text{kg}$-$\text{day}$ has been obtained
after TT and Ped cuts.

\begin{figure}[h]
  \includegraphics[width=1.0\linewidth]{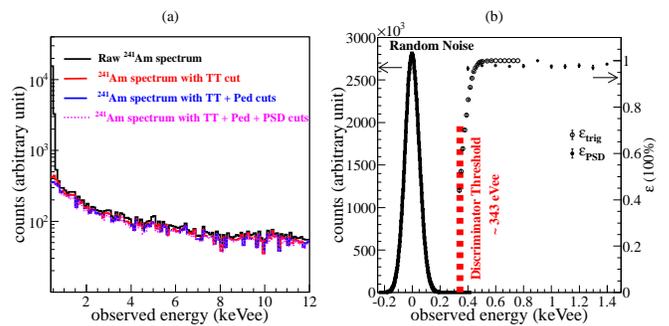}%
  \caption{ \label{fig:r-noise} The $^{241}$Am spectra along the analysis chain
  are shown in Fig.~\ref{fig:r-noise}(a). The distribution of
  the random noise fluctuation and the discriminator threshold of our hardware
  are given in Fig.~\ref{fig:r-noise}(b). The trigger efficiency $\varepsilon_{\text{trig}}$
  and PSD cuts efficiency $\varepsilon_{\text{PSD}}$ have also been shown in the same plot.}
\end{figure}

The second method is the pulse shape discrimination (PSD) cuts which are effective
approaches to reject remaining noise events, including one cut based on the correlation
between the integration area and its maximum amplitude of S$_{\text{p12}}$, and another cut
based on the correlation between signals from p$^+$ electrode and n$^+$ electrode for one event.
The efficiencies of the PSD cuts were measured by a $^{241}$Am source sample and a $^{137}$Cs
source sample ($\varepsilon_{\text{PSD}}$). The spectra of $^{241}$Am source along the analysis chain
are given in Fig.~\ref{fig:r-noise}(a). The efficiencies derived from $^{241}$Am and $^{137}$Cs spectra
agree very well and the bin-by-bin combined $\varepsilon_{\text{PSD}}$ are shown in
Fig.~\ref{fig:r-noise}(b). The source spectra provide consistent measurements of the
TT and Ped cuts efficiencies to those derived from random trigger events.

The distribution of the random noise are given in Fig.~\ref{fig:r-noise}(b).
The energy resolution at this zero energy point is 51 eVee (``ee'' represents electron
equivalent energy). The hardware discrimination threshold, defined at 50\%
trigger efficiency, is calculated about 343 eVee and much higher than the
$5 \cdot \sigma$ (255 eVee). In addition, the trigger efficiency is nearly equal to
100\% above 500 eVee.

\begin{figure}[h]
  \includegraphics[width=1.00\linewidth]{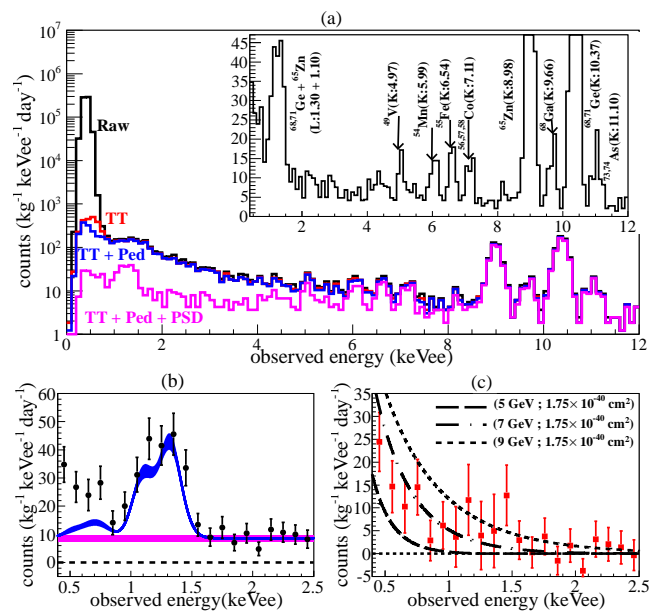}%
  \caption{\label{fig:le-bkg} The observed energy spectra showing raw data, data after TT cut, TT + Ped cuts and
    TT + Ped + PSD cuts were given in Fig.~\ref{fig:le-bkg}(a), respectively. The inset plot in Fig.~\ref{fig:le-bkg}(a) showed
    the background spectrum after TT + Ped + PSD cuts with both the $\varepsilon_{\text{trig}}$,
    cuts efficiencies, fiducial mass, and the dead time correction. Eight $K$-shell peaks
    for $L$-shell peaks prediction are identified.
    The low energy spectrum in the range of 0.4 - 2.4 keVee was shown in Fig.~\ref{fig:le-bkg}
    (b), as well the calculated $L$-shell background contribution and the flat $\gamma$ background
    with the expanded statistical error band. The residual spectrum was shown in Fig.~\ref{fig:le-bkg}(c)
    superimposed with the predicted spectra for 5 GeV, 7 GeV, and 9 GeV WIMP
    with spin-independent cross-section $\sigma_{\chi N} = 1.75 \times 10 ^{-40}$ cm$^2$.}
\end{figure}

Considering both the trigger and signal selection efficiencies, the 400 eVee energy
threshold is selected as our energy threshold for physical analysis.
Fig.~\ref{fig:le-bkg}(a) shows the low energy background spectrum detected by the
PPCGe detector with corrections of the $\varepsilon_{\text{trig}}$ and
cuts efficiencies, fiducial mass and the dead time. Both statistic and systematic
errors are considered with standard error propagation. Several characteristic X-ray
peaks can be seen. They are due to the cosmogenic radioactive isotopes which are mainly
generated within the germanium crystal before installation into CJPL, and include $^{68,71}$Ge, $^{68}$Ga, $^{73,74}$As,
$^{65}$Zn, and so on. Because of the 1.5 mm oxygen-free high-conductivity Copper cryostat surrounding the germanium crystal,
the external low energy x-rays cannot enter into the bulk of the PPCGe detector. Energy calibration was
therefore accomplished by using these internal origin radioactive isotopes. The decays of
the $K$-shell (10.37 keV) and $L$-shell (1.29 keV) peaks of $^{68,71}$Ge
isotopes accord well with their expected half-life \cite{Kang2013}. As the
characteristic x-rays are internal and short-ranged, the detection efficiency is
almost 100\%. The ratios of $K$-shell to $L$-shell X-ray events calculated based on
reference \cite{Bahcall1963} are used to predict the intensity of $L$-shell in
the lower energy ranges ($< 2$ keVee). The background spectrum in the low energy range
of 0.4 - 2.4 keVee and the $L$-shell contributions calculated from the eight clearly visible
$K$-shell peaks has been shown in Fig.~\ref{fig:le-bkg}(b).

We do not apply the surface-bulk cut in this analysis. Accordingly, from simulations and previous
measurements \cite{Aalseth2011,*Aalseth2012,CDMSCollaboration2012,TEXONOCollaboration2013},
there should be a flat $\gamma$ spectrum contributed by the bulk $\gamma$ events which is
mainly located at the internal volume of a PPCGe detector and monotonously decreasing
background spectrum from anomalous surface events due to incomplete charge collection. so that the
expected background should be monotonously decreasing. In addition to the L-shell X-rays contributions,
a conservative flat background level was subtracted at an energy range beyond the tails of the $\chi$-N
nuclear recoil spectrum which, for $m_{\chi} < 12$ GeV, corresponds to 1.7-2.4 keVee.
The final residual spectrum in the region of 0.4-2.4 keVee is shown
in Fig.~\ref{fig:le-bkg}(c) from which the constraints on WIMP are derived.

The thickness of the outer n$^+$ layer of a p-type germanium detector can be measured
by a multi $\gamma$-ray isotope, such as $^{133}$Ba \cite{MAJORANACollaboration2012}.
Due to the close match in total mass and the structure between the CDEX-1 and
TEXONO \cite{Lin2009,TEXONOCollaboration2013} detectors, we chose the same depth of the
dead layer as 1.16 mm with an uncertainty of 10\% for easy comparison.
This gives rise to a fiducial mass of 905 g with an uncertainty
of less than 1\%, corresponding to a data size of 14.6 $\text{kg}$-$\text{day}$.

The quenching factor of the recoiled Ge nucleus is given by the TRIM program
\cite{Ziegler2004,*TEXONOCollaboration2007}. The parameters chosen for the
WIMP in thermal equilibrium includes Maxwellian velocity distribution with
$\nu_0 = 220$ $\text{km}\cdot\text{s}^{-1}$, the escape velocity
$\nu_{\text{esc}} = 544$ $\text{km}\cdot\text{s}^{-1}$ and the local density
($\rho_{\chi}$) of 0.3 $\text{GeV}\cdot\text{cm}^{-3}$. The energy resolution of
the PPCGe was derived from the calibration data and then
extrapolated to the region less than 1 $\text{keV}_{\text{ee}}$.

\begin{figure}[h]
  \includegraphics[width=1.00\linewidth]{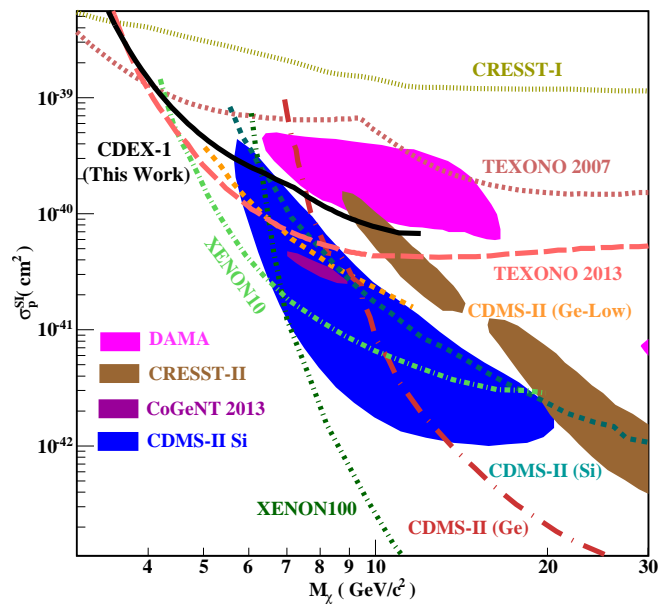}%
  \caption{\label{fig:ex-plot} Exclusion plot of spin-independent coupling,
    superimposed with the results from other benchmark experiments. The results is
    also shown in this figure including the 90\% confidence regions favored by CoGeNT
    \cite{Aalseth2008,*Aalseth2011,*Aalseth2011a,*Aalseth2012}, DAMA/LIBRA \cite{DAMACollaboration2011},
    and CRESST-II \cite{Felizardo2012,*Archambault2012}, as well the exclusion limits from CDMS-II (Si)
    \cite {Ahmed2010,*Ahmed2011,*CDMSCollaboration2012,*Agnese2013a,*Agnese2013b},
    XENON100 and the low-threshold analysis of XENON10 \cite{Aprile2011,*Angle2011,*Aprile2012,*Aprile2012a},
    TEXONO \cite{Lin2009,*TEXONOCollaboration2013}, and CRESST-1 \cite{Angloher2002}.}
\end{figure}

The predicted spectrum of WIMP-nucleon spin-independent interaction can be
evaluated. Using the standard WIMP halo assumption \cite{Donato1998}, the
light-WIMP spectra corresponding to 5 GeV, 7 GeV and 9 GeV WIMP with
spin-independent cross-section $\sigma_{\chi N} = 1.75 \times 10 ^{-40}$ cm$^2$
are also put on the spectrum in Fig.~\ref{fig:le-bkg}(c).

Assuming all of the events from our final residual spectrum are induced by incident WIMP,
we can derive upper limits on the WIMP-nucleon spin-independent cross section at different WIMP masses.
Binned Poisson method \cite{Savage2009} is utilized and the exclusion curve with 90\% C.L.
is displayed in Fig.~\ref{fig:ex-plot}, along with the results from other experiments
\cite{Aalseth2008,*Aalseth2011,*Aalseth2011a,*Aalseth2012,
DAMACollaboration2011, Ahmed2010,*Ahmed2011,*CDMSCollaboration2012,*Agnese2013a,*Agnese2013b,
Angloher2012, Aprile2011,*Angle2011,*Aprile2012,*Aprile2012a, Felizardo2012,*Archambault2012}.
Although we did not apply the bulk-surface cut and the anti-Compton suppression,
the results are close to the latest sensitivities of the TEXONO experiment \cite{Lin2009,TEXONOCollaboration2013}.
The bulk-surface cut could reduce the background level by a factor of 2-3
\cite{Aalseth2008,*Aalseth2011,*Aalseth2011a,*Aalseth2012, Lin2009,*TEXONOCollaboration2013} and
we expect that our new result with bulk-surface cut can be used to check these results.
Also the residual spectrum we have achieved need to be understood further with more data.

An anti-Compton detector will be added to test its performance and background level
in the regime of low cosmic-ray flux. It will need to be evaluated whether the suppression
power of this anti-Compton detector in CJPL can balance its additional contribution to the
PPCGe's background due to its own radioactivities in detail. This will aid in the understanding
of the cosmic-ray induced background when compared with the results from the reference
\cite{TEXONOCollaboration2013}, and it will be very helpful for the
evaluation of the sensitivity of the possible future tonne-scale germanium
experiment in the dark matter search by CDEX collaboration.



\begin{acknowledgments}
  This paper represents the first scientific results obtained at the new underground
  facilities CJPL. The authors would like to thank all those who contributed to its
  efficient construction and commissioning. We are grateful to X. Q. Li and Y. F. Zhou
  for useful comments. This work was supported by the National Natural Science Foundation of China
  (contract numbers: 10935005, 10945002, 11275107, 11175099) and National Basic
  Research program of China (973 Program) (contract number: 2010CB833006).
\end{acknowledgments}

\bibliography{cdex1}

\end{document}